\documentclass[aps,twocolumn,floatfix,prb,showpacs]{revtex4}
\usepackage{epsfig}
\begin{document}
\title{Relaxation kinetics in two-dimensional structures 
} 
\author{Jos\'e L.\ Iguain and Laurent J.\ Lewis }
\affiliation{
D\'epartement de Physique et Groupe de Recherche 
en Physique et
Technologie des Couches Minces (GCM)\\
Universit\'e de Montr\'eal, Case Postale 6128, Succursale Centre-Ville,
Montr\'eal, Qu\'ebec H3C 3J7, Canada.
}
\date{\today}
\begin{abstract}
We have studied the approach to equilibrium of islands and pores
in two dimensions.
The two-regime scenario observed when islands 
evolve according to a set of particular rules, namely relaxation
by steps at low temperature and smooth at high temperature, is generalized
to a wide class of kinetic models and the two kinds of structures.
Scaling laws for equilibration times are 
analytically derived and confirmed by kinetic Monte Carlo simulations.
\end{abstract}
\maketitle

\section{Introduction}
\label{intro}

A large amount of work has been done in the domain
of atomic structure kinetics since 
Burton, Cabrera and Frank presented, in a
seminal paper, the first
serious attempt to model the behavior of atoms adsorbed onto a vicinal
surface \cite{bcf}. Nevertheless, several problems remain unsolved. 
Many important advances  in the techniques for the observation
and manipulation of atoms now
allow dealing with structures of decreasing size and,
with the help of  
powerful computation, 
gaining insight into the basic mechanisms governing atomic dynamics. 
(For a review, see for example Ref.\ \onlinecite{lag,zha,jen0} and references therein).

In the last few years, considerable efforts have been dedicated to understanding 
how nanostructures relax.
A classical description of
structure shape equilibration  was developed
by Herring, Nichols and Mullins (HNM) \cite{mul}. 
In this theory, the transport of mass responsible for relaxation occurs 
in a smooth and continous way via migration
of border adatoms from regions of high curvature (high chemical potential
$\mu$)
 to regions of low curvature (low $\mu$). 
It is assumed that structure 
border is rough enough to allow a continous
(coarse-grained) description of it. However, as this is not valid for surfaces below the
roughening temperature, the low temperature decay  
has been the subject of research both theoretically \cite{jen1,com,tri} 
and experimentally \cite{ex1,ex2}. 

Recently, studies of two-dimensional island shape relaxation with a 
simple model has revealed new, unexpected phenomena \cite{jen1,com}.
It was shown that two
qualitatively different relaxation regimes exist. 
At high temperature, islands evolve toward equilibrium according to HNM.
At low  temperature, islands become  faceted and 
a new driving mechanism appears. It was demonstrated that, in this case,
shape relaxation occurs by steps, where the limiting process consists in the nucleation
of new adatom rows  on the flat edges of islands.
In the above model
adatoms lie on a triangular lattice and both equilibrium
and kinetic properties depend on the single parameter $\epsilon_0/k_BT$
(where $k_B$ is the Boltzmann constant and $T$ the temperature)
since, for an adatom with $n_i$ nearest-neighbors (NN), $n_i\epsilon_0$
is both the potential energy and the kinetic barrier controlling
migration.\par

This two-regime scenario is manifest in the dependence of the equilibration
time $t_{eq}$ on temperature and island size $N$. At high temperature,
\mbox{$t_{eq}\!\sim\! N^2 \exp(3\beta\epsilon_0)$} ($\beta\! =1/k_BT$),
while at low $T$, \mbox{$t_{eq}\!\sim\! N \exp(4\beta\epsilon_0)$}.
The regime to which relaxation of a given island belongs is
determined by border roughness and depends not only on temperature but
also on island size.
In the $T\! - \!N$ plane, the separation line between the two regimes is provided by the 
crossover island size \mbox{$N_c(T)\!\sim\! \exp(\beta\epsilon_0)$},
a rough indication of the size of the largest island that is fully faceted at $T$.  
Interestingly, if the temperature-dependent
factors appearing in the leading terms of $t_{eq}$
are rewritten in terms of $N_c$, the 
properly-scaled equilibration time becomes a function of $N/N_c$ only,
satisfying

\begin{equation}
\left(\frac{t_{eq}}{N_c^5}\right) \sim \left\{
 \begin{array}{ccc}
   \left( \frac{N}{N_c} \right) & \mbox{for} & \frac{N}{N_c} \ll 1 \\
           &      &   \\
   \left( \frac{N}{N_c} \right)^2 & \mbox{for} & \frac{N}{N_c} \gg 1 
 \end{array}
 \right.
\;\;\;\;\mbox{.}
\label{scal0}
\vspace{.3cm}\\
\end{equation}

A first problem of interest is the universality
of this scaling law.
The model described above is at best a 
first approximation to real situations
and many details can be modified
in order to improve it.
For example, such a well-known effect as the Ehrlich-Schwoebel
barrier \cite{schwo} can be accounted for 
by introducing new kinetic barriers in a phenomenological way;
in an even more accurate approach, minimum energy transition paths 
could be calculated from interatomic forces \cite{neb}.  
As a different set of kinetic barriers will entail, to the least, new activation energies,
the question remains whether the above scaling behavior is a consequence of
the model or, to the contrary, would carry over to more general situations.\par

Another closely-related problem is the relaxation of 
 two-dimensional {\it pores}, i.\ e., islands of {\it vacancies}.
In some sense,
islands and pores are `specular images' of one  another.
A given pore has the same boundary  as the corresponding island, and 
so would be faceted or
rough depending on whether the number of {\it vacancies} in it
is smaller or larger than $N_c(T)$. 
It would be desirable to know if pore shape relaxation exhibits also
 a two-regime scenario and to analyze the possibility of 
scaling laws for the corresponding equilibration times.

As far as we know, the questions above have not  been investigated,
even in the case of simple models. 
In this work we study the main properties of 
shape relaxation in two-dimensional structures of adatoms. Our approach is
two-fold: theoretical analysis and numerical simulations.
In the analytic part, we treat the shape relaxation problem 
in a general way. 
Our study suggests the presence of two regimes, regardless of the specific  set of
kinetic barriers and, based on detailed-balance conditions, we show that
similar scaling laws apply for both islands and pores.
Since most of the arguments we invoke are  
heuristic, some independent confirmation of the derived properties is desirable.
This is the goal of the numerical part. 
We analyze, using standard kinetic Monte Carlo (KMC) simulations,
the relaxation of islands and pores according to two models considered
in this paper, and compare the results with our theoretical predictions.\par
 
The paper is organized as follows. In Sec.\ \ref{anal} we discuss the causes for 
scaling behavior and derive the asymptotic laws of scaling functions.
The models we use are defined in Sec.\ \ref{mod} and
the outcome of the KMC simulations are presented in Sec.\ \ref{mc}. Finally, we give our conclusions
in Sec.\ \ref{conclu}.\par

\section{Analytic approach}
\label{anal}

In this section we investigate a
generic model with adatoms lying on a triangular
lattice where 
the total binding energy
per NN pair is $\epsilon_0$. Equilibrium properties  depend  on 
a single parameter, namely $\beta\epsilon_0$, but
kinetics  involve in general
a greater number of them. Our aim here is to establish 
 scaling laws  for both island and pore relaxation times.

Let us consider islands first. 
It is easy to see that many of the arguments presented in Ref.\ \onlinecite{jen1,com}
remain applicable regardless of the specific set of kinetic barriers.
First of all, $N_c(T)$, which separates rough and faceted islands at a given temperature,
is the size at which the perimeter equals the average distance between
border kinks. This does not depend on the transition barriers but only on
the binding energies in the different configurations, so the form 
$N_c\sim\exp(\beta\epsilon_0)$ is still valid.
Next, two qualitatively different modes of relaxation should exist.
On the one hand, for islands greater that $N_c$, shape relaxation dynamics 
 is adequately described by the HNM theory. It consists of a set of equations
for the temporal evolution of the border curvature, where the kinetics is
dominated by perimeter diffusion and always leads to a $t_{eq}\!\sim\!\! N^2$
power law. We will call this regime {\it rough relaxation mode} (RRM).
On the other hand,   
mechanisms driving islands smaller than $N_c$ toward equilibrium are also quite general.
In this scenario, islands spend most of the time in highly-faceted configurations.
We will call this regime {\it faceted relaxation mode} (FRM).
Evolution occurs by steps consisting in nucleation and stabilization of
new adatom rows.  
The time associated to
each of these steps is proportional to facet length
($\sim\!\! N^{1/2}$) and,
given an initial configuration, a number $\sim\!\! N^{1/2}$ of rows
needs to be created in order to attain the equilibrium shape; thus $t_{eq}\!\sim\!N$.

The size exponents for equilibration time thus
appear to be universal (1 in the FRM, 2 in the RRM).
The activation energies however 
{\it do} depend in general 
on the kinetic barriers. The activation energy in the FRM corresponds to 
the energy characteristic of row nucleation, while in the RRM it will be equivalent to the
characteristic energy for diffusion along the island border. 
 If we call these energies $E_F$ and $E_R$ for the FRM and RRM, respectively, 
the arguments presented so far allow us to write the ansatz

\begin{equation}
t_{eq} \sim \left\{
 \begin{array}{ccc}
   \left( \frac{N}{N_c} \right) e^{\beta(E_F+\epsilon_0)} & \mbox{for} & \frac{N}{N_c} \ll 1 \\
                   &            &         \\
   \left( \frac{N}{N_c} \right)^2 e^{\beta(E_R+2\epsilon_0)} & \mbox{for} & \frac{N}{N_c} \gg 1
 \end{array}
 \right.
\;\;\;\;\mbox{.}
\label{gene}
\vspace*{.5cm}\\
\end{equation}

At this point, it is clear that the occurrence of scaling is limited to the cases where 

\begin{equation}
E_F=E_R+\epsilon_0\;\;\;\;\mbox{;}
\label{rel}
\end{equation} 

\noindent
otherwise, the exponential factors in Eq.\ (\ref{gene})
would be different and no {\it natural} time scale [such as $\sim\!\! N^5$ in Eq.\ (\ref{scal0})],
allowing scaled $t_{eq}$ to be expressed as a function of $N/N_c$, would exist.
Let us, then, analyze the connexion between the two energies. 

\begin{figure}[ht]
\centerline{
\psfig{file=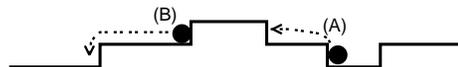,width=2.5cm,angle=-90,clip=}
}
\caption{Schematic of the limiting processes for diffusion along a rough border.
Adatom detaches from a kink and go into adjacent rows after jumping around
a corner. All the sites below the full line, which represents the border, are occupied.
}
\label{broken}
\end{figure}

As noted above, the limiting step in the RRM 
is adatom border diffusion.
This is
in turn limited by the elementary processes sketched in Fig.\ \ref{broken},
in which adatoms pass from one kink to another by 
jumping around a corner; thus we can readily equal $E_R$ and the energy of this process. 
 
For the FRM, the dynamics is determined by  row nucleation, which occurs
when two adatoms meet on a flat island edge.
The nucleation rate may thus be calculated by the product of two factors:
 the rate of adatoms entering a facet
and the probability that another adatom lies on the same facet.
In this regime,
the main sources of adatoms are facet ends, where the most weakly bound adatoms are found.
 Those adatoms  move onto flat facets by basically the 
process sketched in Fig.\ \ref{broken}\ (A);
 the first factor is thus
\mbox{$\sim \exp(-\beta E_{R})$}. The second factor is 
$\sim \exp(-\beta\epsilon_0)$ because the system gains
an energy $\epsilon_0$ when an adatom
moves from a kink to a flat edge. Thus, we have
$E_F=E_R+\epsilon_0$, i.\ e., Eq.\ (\ref{rel}), and  
the scaling form

\noindent

\begin{equation}
\frac{t_{eq}}{N_c^\alpha} \sim \left\{
 \begin{array}{ccc}
   \left( \frac{N}{N_c} \right) & \mbox{for} & \frac{N}{N_c} \ll 1 \\
           &      &   \\
   \left( \frac{N}{N_c} \right)^2 & \mbox{for} & \frac{N}{N_c} \gg 1
 \end{array}
 \right.
\;\;\;\;\;\mbox{,}
\label{scal1}
\vspace*{.5cm}\\
\end{equation}

\noindent
where the exponent $\alpha$ is 

\begin{equation}
 \alpha=2+\frac{E_{R}}{\epsilon_0}\;\;\;\;\;\mbox{.}
\vspace*{.3cm}\\
\label{alpha}
\end{equation}

We consider now the case of pores. Since, as mentioned before, a pore will also be rough or faceted 
depending on whether it is larger 
 or smaller than $N_c$, it is worth analyzing 
each situation separately. 

For the relaxation of rough pores, the solution is immediate: 
 the kinetic path    
is the same as that followed by the corresponding island (via a vacancy-atom exchange mechanism).
This pore-island symmetry is clearly evident in the HNM theory 
(see for example Ref.\ \onlinecite{jen1}), 
which involves only border curvature and perimeter atomic diffusion
and hence does not depend on the kind of particles enclosed within 
the border. As a corollary, the equilibration time for pores should also satisfy
the asymptotic law established by the second equation in Eq.\ (\ref{gene}); this
will be verified numerically in Sec.\ \ref{mc}.

\begin{figure}[h]
\centerline{
\psfig{file=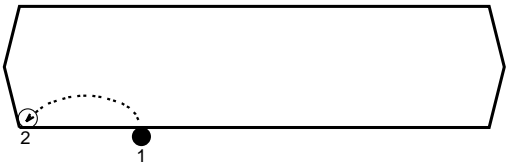,width=2.5cm,angle=-90,clip=}
}
\caption{Schematic of row nucleation in a pore. All adatoms are outside the
line which represents the border of a fully faceted pore.
}
\label{hexa}
\end{figure}

For faceted pores, it is clear that relaxation will occur, as in the case of faceted
islands, by steps consisting of the nucleation and
stabilization of new rows. However, despite this qualitative analogy, 
the two processes are not equivalent but, rather, complementary;  adatom
migration occurs from short to long facets in islands, while the opposite
is true in pores. For the latter, row nucleation  
occurs when, as  shown schematically in Fig.\ \ref{hexa}, 
an adatom that lies on a long facet moves to an internal corner of the pore.
The new row will grow with the arrival of
more adatoms from the now `open' long facet, and become stable when fully filled. 

The similarity of the relaxation mechanisms allows us to apply, in
the case of pores, the same kind of arguments we invoked for 
the dependence of $t_{eq}$ on the size of faceted
islands, which again yields the power law $t_{eq}\sim\! N$.
In order to calculate the activation energy --which corresponds
to the nucleation    
process illustrated in Fig.\ \ref{hexa}--
 we consider first the inverse process, i.\ e.,
the migration of an adatom from site $\bf 2$ to site $\bf 1$. 
In this case, the adatom first leaves a kink, then diffuses along 
a flat border and, finally, goes into an adjacent row by jumping around a corner. 
This is similar to the process sketched in Fig.\ \ref{broken}\ (B) and 
must therefore have an activation energy $E_R$.
Knowledge of this last energy allows us to find that correponding to 
nucleation. 
They are related by the detailed-balance condition which, taking into accout
that the binding energy difference between sites $\bf 1$ and $\bf 2$ is $\epsilon_0$,
finally leads to the result that the activation energies for pore shape relaxation 
are also connected by  relation (\ref{rel}). 

As an interesting consequence, the 
equilibration time for islands or pores, when properly
scaled with $N_c^\alpha$, becomes a function of $N/N_c$ only that
satisfies the asymptotic rules (\ref{scal1}).
We stress that even though the HNM theory implies 
an equivalent relaxation mode for both kinds of structures
in the RRM, the same symmetry
cannot be expected to hold in the FRM. The asymptotic forms 
give  the exponents of the leading terms 
but different prefactors may appear. In general,
there will be one scaling function corresponding to islands 
and {\it another} scaling function for pores. For 
very small values of their arguments ($N\! \ll\! N_c$) they will be
parallel (on a log scale) but not necessarily equal, while at the other extreme ($N\! \gg\! N_c$) 
the curves should collapse as required by HNM theory.
Using KMC we will see, in Sec.\ \ref{mc}, that these properties of the
scaling functions are verified.

\section{The Models}
\label{mod}

In order to asses the analytical predictions above,
we investigated two different models, denoted  I and II, which have the
following common characteristics:  
The substrate is represented by a triangular lattice and
the total binding energy per
NN pair is $\epsilon_0$.
Adatom hoppings are only possible between NN sites and 
a given adatom in an initial site $a$ jumps to a final empty  site $b$
with a probability per unit time \mbox{$P_{ab}=\nu \exp(-\beta E_{ab})$}, 
where $\nu$ is a constant
frequency and $E_{ab}$ the  kinetic barrier. 
In order to avoid detachment, the jumps are  
forbidden if the number of NN adatoms in the final site is 0.  
As an additional simplification, the motion of adatoms with initially more
than 4 NN is also forbidden. This approximation is
justified because of the high energy
barriers of the corresponding processes; it is introduced
in order to accelerate the simulations. 
Differences between models I and II lie only in the set of kinetic barriers.

For Model I, which is the same as the one used in Ref.\ \onlinecite{jen1,com},
 $E_{ab}$ depends only on the number $n_a$ of NN adatoms
in the initial site $a$. It takes the value \mbox{$E_{ab}=n_a \epsilon_0$}
 when \mbox{$1\! <\! n_a\! <\! 5$} and
\mbox{$E_{ab}=\epsilon/10$} when \mbox{$n_a\! =\! 1$}, 
regardless of the number of NN adatoms in $b$.

\begin{figure}[ht]
\centerline{
\psfig{file=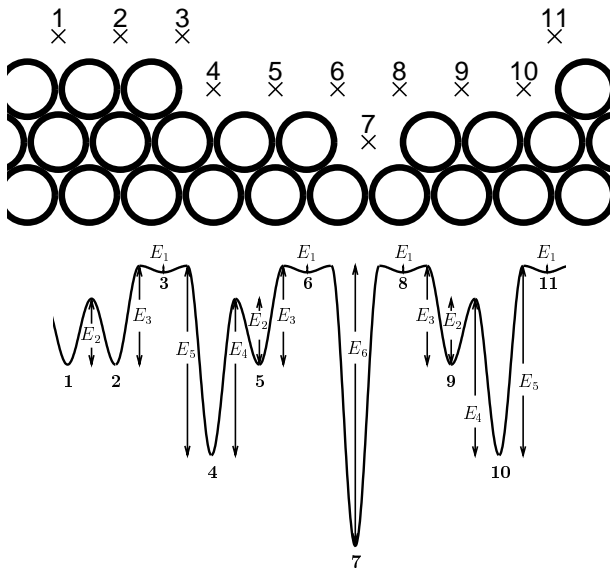,width=8.cm,clip=}
}
\caption{
The energy landscape for Model II.
}
\label{modeloII}
\end{figure}

For Model II, we use a more accurate set of activation energies, obtained from a
 detailed study of the kinetic barriers for adatoms
lying on a triangular lattice and interacting via a Lennard-Jones potential 
 \cite{thibault}.       
 It is useful to define this model with the help of Fig.\ \ref{modeloII}, where the different
 barriers are represented. 
In the top part of the figure, adatoms are represented by circles while crosses indicate possible 
binding sites. The bottom part shows the corresponding energy landscape, i.\ e., kinetic barriers 
separating adjacent local minima.
Thus, in this model, we take into account the final-state configuration when evaluating
the transition-state energy. (Note however that the height of the barrier depends on
the number of NN in the initial and final states but not on their precise position).
Model II thus incorporates such important features as the Ehrlich-Schwoebel  barrier.
This is apparent if we look at Fig.\ \ref{modeloII}:  the barrier  from
$\bf 2$ to $\bf 1$ is lower than the barrier from $\bf 2$ to $\bf 3$; the barrier from
$\bf 4$ to $\bf 3$ is higher than the barrier from $\bf 4$ to $\bf 5$, etc.
 The final parameters in this model  are
\mbox{$E_1\! =\! 0$}, \mbox{$E_2\! =\! 0.5\epsilon_0$}, \mbox{$E_3\! =\! \epsilon_0$},
\mbox{$E_4\! =\! 1.5\epsilon_0$}, \mbox{$E_5\! =\! 2\epsilon_0$},
\mbox{$E_6\! =\! 4\epsilon_0$}.
In order to facilitate comparison, Table
\ref{table} lists, for both models, the kinetic barriers corresponding to a given elementary jump
as a function of the number of NN  before ($n_i$)
 and after ($n_f$) the hopping.

\begin{table}[h]
\centerline{
\begin{tabular}{|r|c|c|}\hline
$n_i$ & $\Delta\!E^{I}$[$\epsilon_0$]  & $\Delta\!E^{II}$[$\epsilon_0$]\\ \hline\hline
1     &   0.1           &  0.0               \\ \hline
      &                 & 0.5, for $n_f\!\ne\!1$ \\ 
2     &   2.0             &                   \\ 
      &                 &   1.0,  for $n_f\!=\!1$  \\ \hline
      &                 &  1.5, for $n_f\!\ne\!1$ \\
3     &   3.0             &                      \\
      &                 &   2.0, for $n_f\!=\!1$  \\ \hline
4     &   4.0             &   4.0              \\ \hline
\end{tabular}
}
\caption{Kinetic barrier for Model I ($\Delta\!E^{I}$) and Model II ($\Delta\!E^{II}$)
as a function of the number of NN before ($n_i$) and after ($n_f$) the jump.
For Model I, the barriers depend only on $n_i$.
Detachments, as well as jumps of adatoms with $n_i\! >\! 4$, are forbidden in both models.
} 
\label{table}
\end{table}

Before closing this section, we calculate the activation energies 
 for each model as derived from the analysis of 
Sec.\ \ref{anal}. Remember that $E_R$ is the energy characteristic
of the process sketched in Fig.\ \ref{broken}\ (A).
For Model I, this process corresponds to an adatom with 3 NN before the jump,  so 
 \mbox{$E_R^{I}\! =\! 3\epsilon_0$}, and, using Eqs.\ (\ref{gene}) and (\ref{alpha}),
 \mbox{$E_F^{I}\! =\! 4\epsilon_0$} and \mbox{$\alpha^{I}\! =\! 5$}, as obtained
in Ref.\ \onlinecite{jen1,com} (but not yet tested with pores).
For Model II, the process in question is (for example) the transition from $\bf 4$ to  $\bf 2$ 
in Fig.\ \ref{modeloII}, which leads to 
 \mbox{$E_R^{II}\! =\! 2\epsilon_0$},
{$E_F^{II}\! =\! 3\epsilon_0$} and \mbox{$\alpha^{II}\! =\! 4$}.

\section{Numerical results}
\label{mc}

As mentioned in Sec.\ \ref{intro}, a numerical confirmation of the 
results of Sec.\ \ref{anal}, obtained in a heuristic manner, is in order. 
In this section, we do this using KMC simulations to 
explore shape relaxation according to models I and II.
We consider islands and
pores with sizes between 400 and 6500 particles and perform, for each model, 
standard KMC simulations.
The values of $\beta$ (in $1/\epsilon_0$ units) range between 2 and 17. 
The results shown in this section correspond to averages over 
a number of samples, varying from 4 for the largest structures to
30 for the smallest ones.

Following Ref.\ \onlinecite{jen1,com}, we use the aspect ratio $\alpha$, defined as the ratio of
the $x$ and $y$  gyration radii, to characterize the  state of
the structures. 
 We start each simulation with an aspect ratio around 10 and
define the equilibration time $t_{eq}$ as the  time at which
$\alpha$ first becomes less than 1.
First of all, we analyze the dependence of $t_{eq}$ on $N$ (for fixed $\beta$) and,
separately, on $\beta$ (for fixed $N$). Later in this section, we will check the
scaling laws.

\begin{figure}[h]
\centerline{
\psfig{file=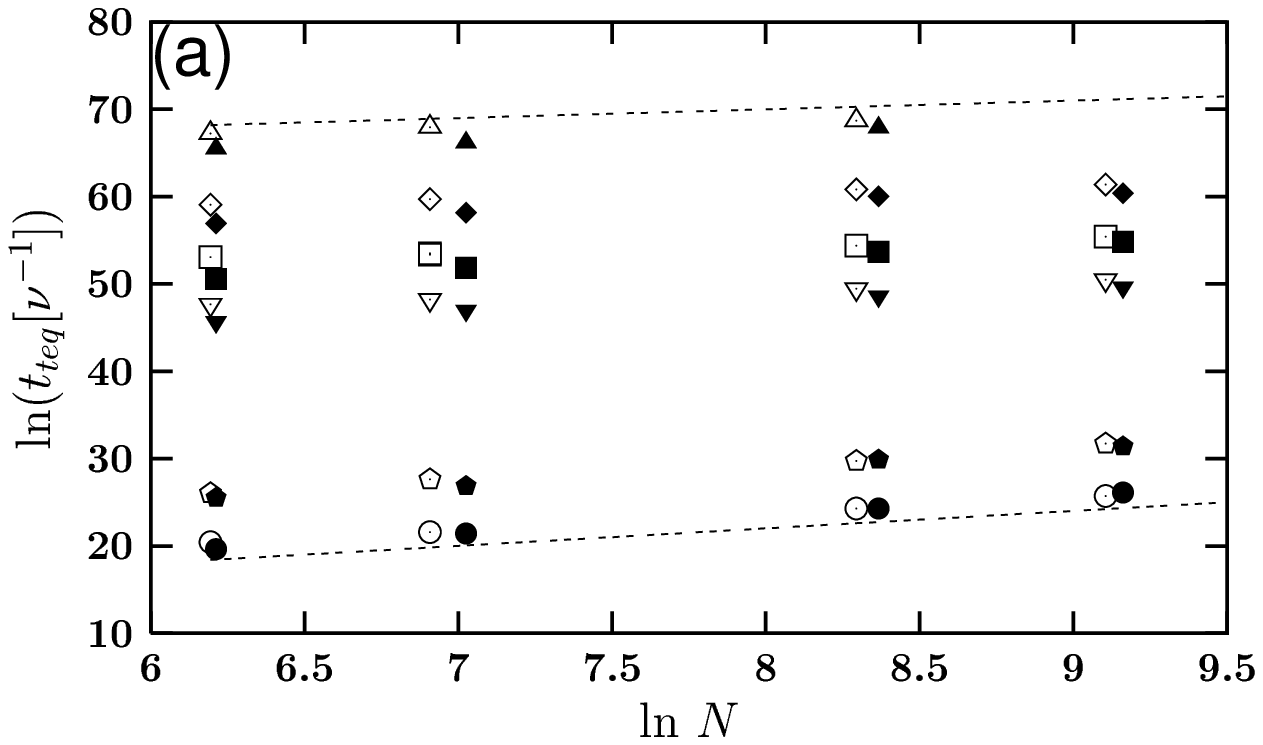,width=8.cm,clip=}
}
\centerline{
\psfig{file=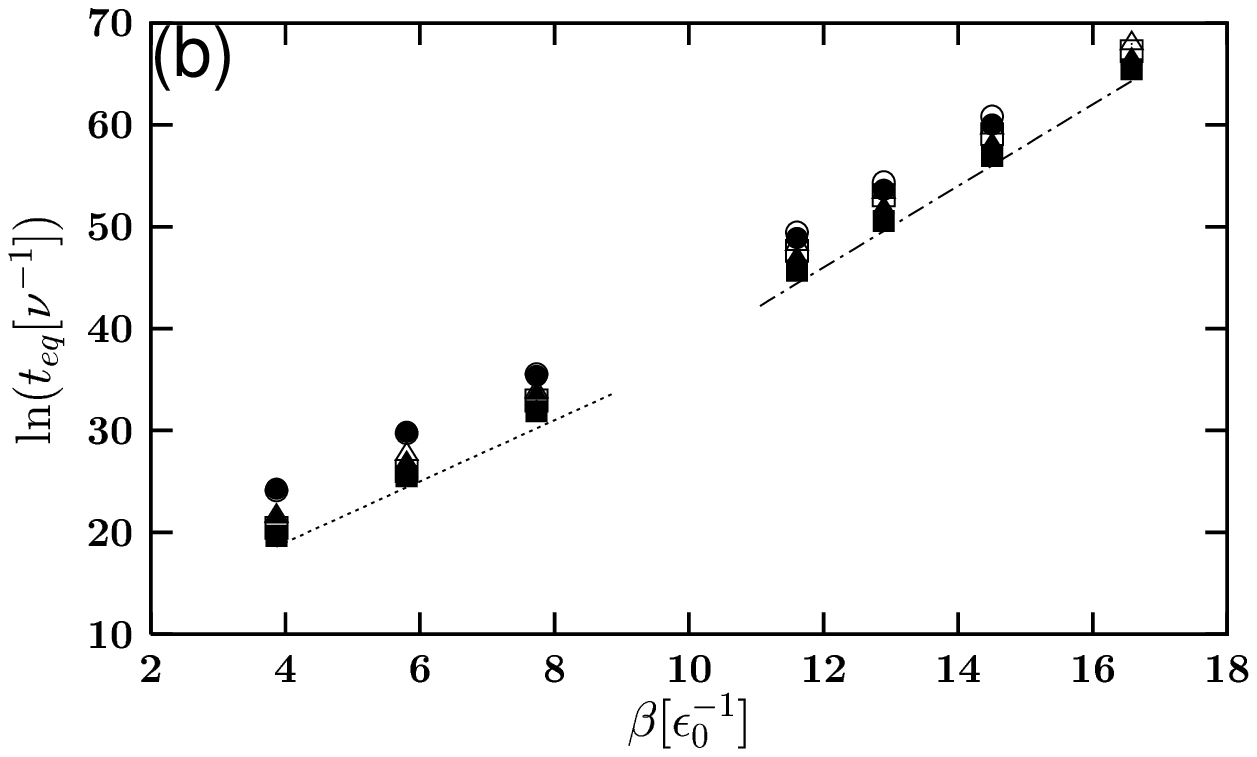,width=8.cm,clip=}
}
\caption{Equilibration time for Model I.
(a) $t_{eq}$ as a function of $N$ for $\beta=$ 3.9 (circles), 5.8 (pentagons), 11.6 (down-triangles),
12.9 (squares), 14.5 (diamonds), 16.6 (up-triangles); the lines have
slopes of 1 (upper) and 2 (lower).
(b) $t_{eq}$ against $\beta$ for
$N=$ 490 (squares), 1000 (triangles), 4000 (circles);
the lines have slopes of 3 (left) and 4 (right).
Filled symbols correspond to islands and empty ones to pores.
}
\label{simple-lim}
\end{figure}

Let us start with Model I. In order to test the behavior of $t_{eq}$ as a function
of $N$, we plot in 
 Fig.\ \ref{simple-lim}\ ($a$) 
 $\ln t_{eq}$
against $\ln N$ for several temperatures. Filled and empty symbols correspond to 
islands and pores, respectively.
At high (low) enough temperature, $t_{eq}$ is expected to scale as $N^2$ ($N$).
This scaling behavior is represented in Fig.\ \ref{simple-lim}\ ($a$)
by the lines with slopes of 2 (lower) and 1 (upper), which bracket
the KMC results corresponding to the highest and lowest
temperatures used in our simulations, respectively.   
Using these lines as a guide, it is indeed clear that the 
the size exponent changes from 1 to 2
when the temperature is increased, consistent with the numerical findings.
Note that the agreement
is as good for pores as it is for islands.\par
 
We check now our predictions for the activation energies.
To do so, in Fig.\ \ref{simple-lim}\ ($b$),  $\ln t_{eq}$
is plotted against $\beta$ for different sizes.
The asymptotic  behaviors are represented
by the straight lines which have slopes of $3\epsilon_0$ (low $\beta$)
and $4\epsilon_0$ (high $\beta$). 
The KMC data again show excellent agreement with the 
predicted behavior, to within a trivial prefactor (i.\ e., slopes agree).
Let us remark that the results for island relaxation are given here
to allow comparison with the (new) results for pores.
We note however a small discrepancy between our numerical results and those
of Ref.\ \onlinecite{com}: although the size exponents and activation energies coincide,
our equilibration times are found to be lower (by a factor of $\sim\! 10$)
than those reported in Ref. \onlinecite{com}.

\begin{figure}[h]
\centerline{
\psfig{file=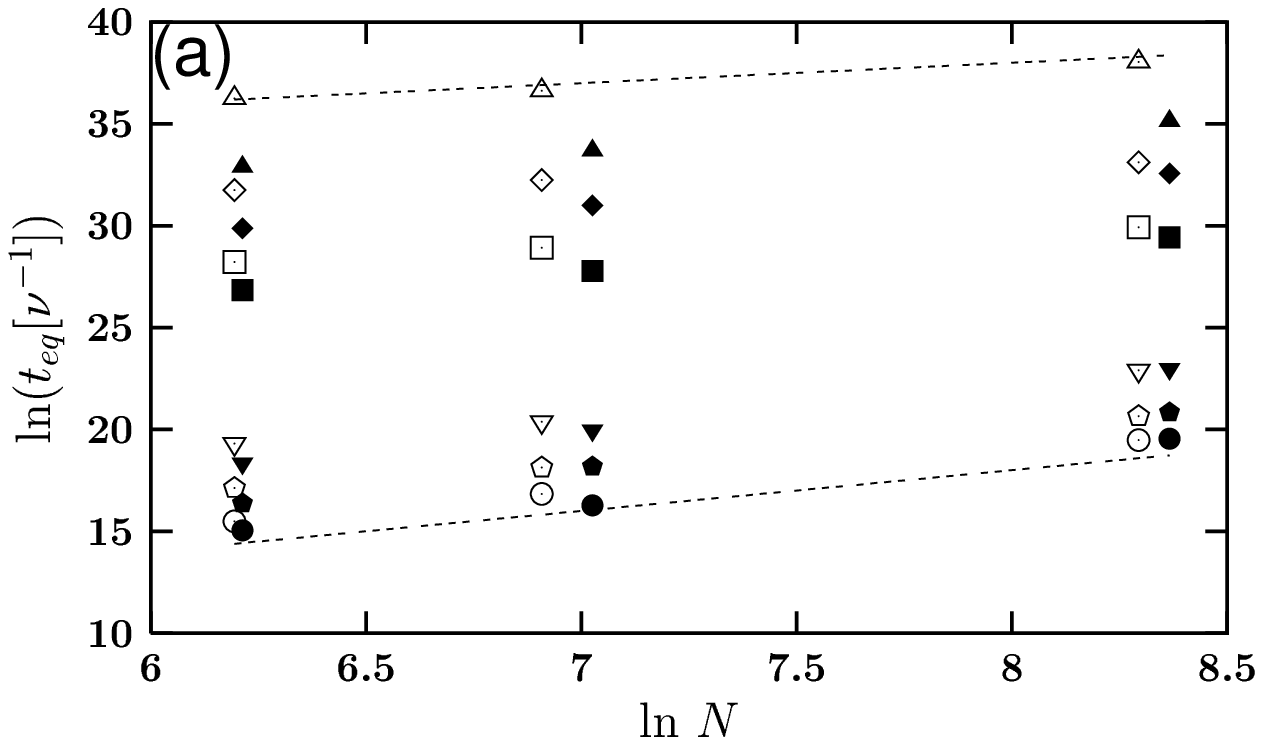,width=8.cm,clip=}
}
\centerline{
\psfig{file=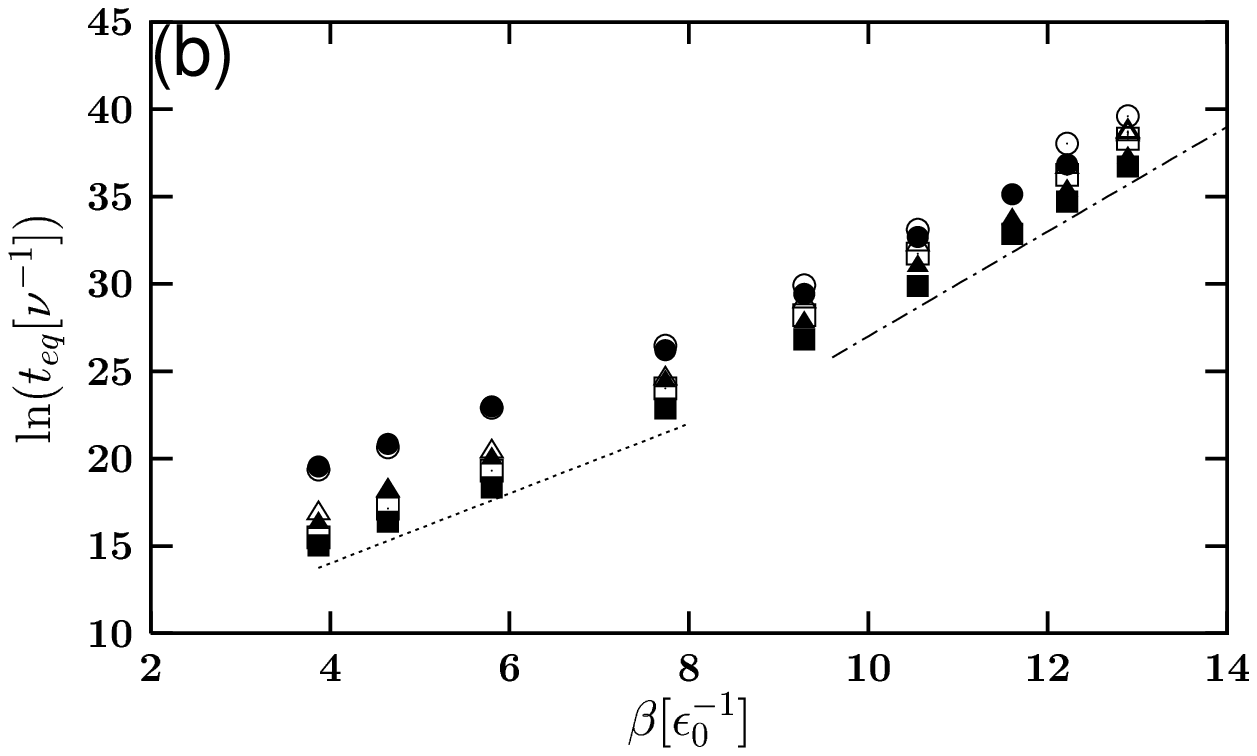,width=8.cm,clip=}
}
\caption{Equilibration time for Model II. (a) $t_{eq}$ as a function 
of $N$ for $\beta=$ 
3.9 (circles), 4.6 (pentagons), 5.8 (down-triangles),
10.6 (squares), 11.6 (diamonds), 12.2 (up-triangles);
the lines have slopes of 1 (upper) and 2 (lower).
(b) $t_{eq}$ against $\beta$ for $N=$ 490 (squares), 
1000 (triangles), 4000 (circles);
the lines have slopes of 2 (left) and 3 (right). 
Filled symbols
correspond to islands and empty ones to pores.
}
\label{schwo-lim}
\end{figure}

The corresponding results for Model II are presented in Fig.\ \ref{schwo-lim}.
Note that, also in this case, the theory-simulation agreement is good for both
types of structures.
Thus, Figs.\ \ref{simple-lim} and \ref{schwo-lim}  
confirm our predictions that, on the one hand,
the size exponents are universal (1 and 2) and, on the
other hand, $E_R$ and $E_F$ depend on the specific set of microscopic kinetic barriers.

\begin{figure}[ht]
\centerline{
\psfig{file=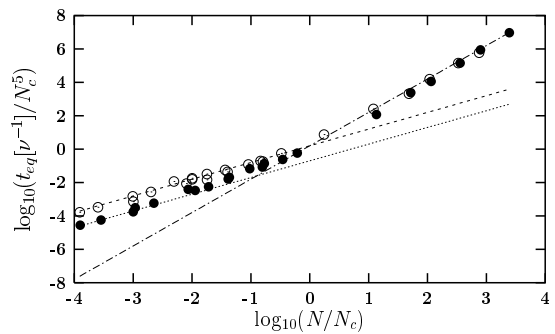,width=8.cm,clip=}
}
\caption{
Model I scaling of $\log_{10}(t_{eq}/N_c^5)$ against $\log_{10}(N/N_c)$,
for islands (filled symbols) and pores 
(empty symbols). The lines have slopes of $1$ and $2$.
}
\label{simple-sca}
\end{figure}

\begin{figure}[ht]
\centerline{
\psfig{file=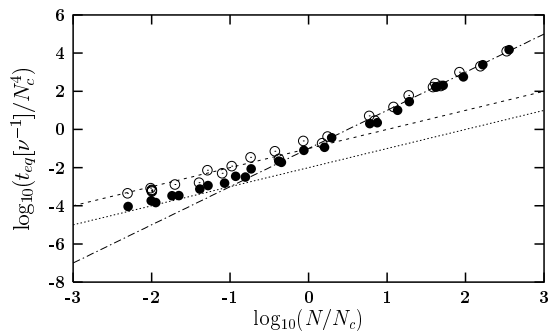,width=8.cm,clip=}
}
\caption{Model II collapse of data
for both islands (filled symbols) and  pores
(empty symbols). The lines have slopes of $1$ and $2$
}
\label{schwo-sca}
\end{figure}

Finally, in order to test the specific scaling laws,
we plot, on a log-log scale, $(t_{eq}/Nc^5)$
against $(N/Nc)$ for Model I (Fig.\ \ref{simple-sca}) and $(t_{eq}/Nc^4)$
against $(N/Nc)$ for Model II (Fig.\ \ref{schwo-sca}). 
We have used $\alpha^I$ and $\alpha^{II}$  as calculated 
at the end of Sec.\ \ref{mod}. 
According to the analysis presented in Sec.\ \ref{anal}, $t_{eq}/N_c^{\alpha}$ should scale
with exponents 1 and 2 for very small and very large values of its argument ($N/N_c$). 
In Figs.\ \ref{simple-sca} and \ref{schwo-sca}, we have  drawn lines with slopes 
of 1 and 2 to indicate  these asymptotic behaviors, which are evidently closely
verified.
For concreteness, in these plots, we use the same \mbox{$N_c\! =\! .25\exp(\beta\epsilon_0)$} as in
Ref. \onlinecite{jen1}; the exact value of the geometrical prefactor should not affect  
the final result in a significant manner. 

It is interesting to note that, besides confirming  the expected asymptotic power laws,
in both figures all the points corresponding to islands collapse on one curve,
and all the points corresponding to pores collapse on another one (depending via $\alpha$ on 
the microscopic details of the model), 
thus giving additional support to the scaling ansatz.
Furthermore, although for each model the island and
pore scaling functions are different in the FRM, they become equivalent
in the RRM, as required by the perfect pore-island symmetry of relaxation in the
latter regime.

\section{Conclusions}
\label{conclu}

By considering a generic model of 
adatoms lying on  a triangular lattice, we have
studied the problem of shape relaxation of islands and pores.
The arguments employed in this work
allows 
the main properties of the equilibration time
$t_{eq}$ to be calculated as a function of temperature and  size.
Because our arguments are somewhat heuristic,
 KMC simulations, using two different kinetic models, 
were also carried out. The numerical results confirm our theoretical predictions. 
For both islands and pores
 two qualitatively different modes
of relaxation (FRM and RRM) are found,  as well
as the line $N_c(T)$ that separates these regimes in 
the temperature-size plane. It was shown that, although size exponents
are universal (1 in FRM, 2 in RRM), the activation energies corresponding
to each mode depend on the microscopic details of the kinetic model. 
Scaling behaviors were found for the equilibration time: 
the properly scaled $t_{eq}$ depends also, via $\alpha$, on
the set of elementary kinetic barriers of the model.  
Furthermore, for a given model, the specific scaling function for 
islands is in general different from the one corresponding to 
pores. While in the FRM both functions scale with the same exponent ($=\!\! 1$)
 as a consequence of the detailed-balance condition, in the
RRM the two functions collapse because of the perfect island-pore
symmetry of  HNM theory \cite{mul}.\vspace{.5cm}\\

\section*{AKNOWLEDGMENTS}
We are grateful to N.\ Combe and P.\ Jensen for interesting discussions.
We thank F.\ Bussi\`eres, M. de la Chevroti\`ere, J.\ Richer, M.\ Simard and 
C.\ Hudon for help with the numerical calculations, 
and P.\ Thibault who provided the data needed for  Model II.
This work was supported by grants from the
Natural Sciences and Engineering Research Council (NSERC) of Canada and
the ``Fonds Qu{\'e}b{\'e}cois de la
recherche sur la nature et les technologies'' (FQRNT) of the Province of
Qu{\'e}bec. We are indebted to the ``R\'eseau qu\'eb\'ecois de calcul de
haute performance'' (RQCHP) for generous allocations of computer
resources.

\newpage

\end{document}